\documentclass[11pt]{article}
\usepackage{deauthor,times,graphicx}
\usepackage{url}
\usepackage{xspace}
\usepackage{wrapfig}
\usepackage{colortbl}
\usepackage{multirow}
\usepackage{epsfig}

\graphicspath{{./}}
\newcommand{\hidden}[1]{}
\newcommand{\empar}[1]{\vspace{3pt}\noindent{\em #1}}
\newcommand{\sampar}[1]{\smallskip\noindent{\bf #1}}
\newcommand{\stitle}[1]{\vspace{3pt}\noindent{\bf #1}}
\newcommand{\emtitle}[1]{\vspace{3pt}\noindent{\bf #1}}
\newcommand{\mar}{{\bf MARQED}\xspace}
\begin{document}

\title{Optimizing Open-Ended Crowdsourcing: The Next Frontier in Crowdsourced Data Management}
\author{Aditya Parameswaran$^\dagger$, Akash Das Sarma$^\star$, Vipul Venkataraman$^\dagger$ \\
$^\dagger$University of Illinois \ \ \ \ $^\star$Stanford University
}
\maketitle

\begin{abstract}
Crowdsourcing is the {\em primary means to generate training data at scale}, 
and when combined with 
sophisticated machine learning algorithms, 
crowdsourcing is an  enabler for a variety of emergent automated applications
impacting all spheres of our lives.  
This paper surveys the emerging field of 
formally reasoning about and optimizing {\em open-ended crowdsourcing},
a {\em popular and crucially important, 
but severely understudied} class of crowdsourcing---the 
next frontier in crowdsourced data management. 
The underlying challenges include 
distilling the right answer when none of the workers agree with each other,
teasing apart the various perspectives adopted by workers when answering tasks, and effectively selecting between the many open-ended operators appropriate
for a problem. 
We describe the approaches that we've found to be effective for open-ended
crowdsourcing, drawing from our experiences in this space.
\end{abstract}

\section{Introduction}

\label{sec:intro}

\noindent We are on the cusp of a new data-enabled era, where machine learning
algorithms, trained on large volumes of labeled training data, are enabling increasing 
automation in our daily lives---from driving, robotics, and manufacturing, to surveillance, medicine, and science;
a recent New York Times article calls this {\em ``a transformation that many believe will have a payoff on the scale of the personal computing industry or the commercial internet''}~\cite{ai-nytimes}. 
Although considerable effort has gone into the development of 
machine learning algorithms for these applications, the generation of labeled training datasets,
done at scale using {\em crowdsourcing}~\cite{crowd-book}---
while equally important~\cite{halevy2009unreasonable,domingos2012few}---is often overlooked.
Recent work has demonstrated that {\em optimizing crowdsourcing} can often yield
{\em orders of magnitude more high-quality labeled data at the same cost}, spurring 
the development of increasingly sophisticated machine learning algorithms,
and providing immediate benefits via substantial increases in accuracy for existing algorithms. 

\sampar{From Boolean Crowdsourcing to Open-Ended Crowdsourcing.} 
However, most of the past research on optimizing crowdsourcing 
has focused on what we call 
{\em boolean crowdsourcing}, e.g.,~\cite{crowdscreen, rate, markus-sorts-joins, searching, DBLP:conf/webdb/PolychronopoulosADGP13,crowder,whowon}, 
where human involvement can be abstracted as tasks or operators 
whose answers come from a small, finite domain,
e.g., evaluating a predicate, comparing a pair of items, or rating an item on a scale from 1--5.
In the former two cases, the domain of possible answers is boolean, while in the
last case, the domain is finite and has cardinality five. 
Boolean crowdsourcing operators have natural analogs in computer 
operators, and are easier to reason about and develop algorithms with.
This research has largely ignored {\em open-ended crowdsourcing}, where 
 the answers to the tasks have no similar restriction.
Anecdotal evidence indicates that open-ended crowdsourcing
tasks are at least as popular as traditional boolean crowdsourcing tasks
on crowdsourcing marketplaces.
As one data point, an analysis of Mechanical Turk's log data~\cite{difallah2015dynamics}
reported that {\em content creation}---including open-ended tasks like
transcription---was by far the single largest category of tasks in the period from 2009--14. As another example, a survey of industry users of crowdsourcing reports similar findings~\cite{crowd-book}.

\sampar{Examples of Open-Ended Crowdsourcing.}
We now describe some canonical open-ended crowdsourcing problems
that we use as examples for this paper. 
One example is {\em transcription}: the goal of transcription is 
to transform a piece of audio or video to text. 
This could be done,
for example, 
using unit tasks where 
we provide workers a portion of the audio or video, along with 
a text box where they can type a sequence 
of words to match the contents of the portion.
Another example is {\em clustering}: the goal of clustering is
to subdivide a collection of items (images, pieces of text) into clusters.
This could be done, for example, using unit tasks where workers
place items into an arbitrary number of buckets determined by them.
Yet another example is {\em detection}: the goal of detection
is to identify where in an image an object is located.
This could be done, for example, using unit tasks where workers
draw a bounding rectangle or polygon around the location of
the object of interest.

\sampar{Open-Ended Crowdsourcing: A Pressing Need.} 
Beyond the popularity, there are several reasons why 
open-ended crowdsourcing is crucially important.
First, some tasks are {\em near-impossible with 
just boolean crowdsourcing}.
For example, using boolean crowdsourcing to locate the 
position of an object in an image is near-impossible.
If we use a task like: ``is the object in this portion of the image (yes/no)'',
we may be able to get close to the actual location of the object by repeatedly
using this task on various portions of the image, but
getting an accurate bounding box around the object may require 
hundreds or thousands of such tasks.
On the other hand, asking workers to simply draw a bounding box
is a lot more effective in terms of time, cost, and accuracy.
Second, open-ended crowdsourcing lets us {\em get more fine-grained data},
since workers provide answers from a potentially unbounded set.
If we were to use an entropy argument, the number of bits provided by workers for 
an open-ended question or task 
is considerably larger than that provided for a boolean question. 
Third, recent computer vision and text processing papers
argue that {\em fine-grained training data is essential} for developing sophisticated
machine learning models~\cite{fine-nlp1,fine-nlp2,fine-cv1,bell2014intrinsic}. 
Specifically, our best hope for improving the accuracy of 
present machine learning models is by using training data 
that reveals more information about how humans think, as 
opposed to training data that is more akin to how computers
operate (i.e., via boolean operations).

{\em Despite these compelling reasons, 
research on optimizing open-ended crowdsourcing is still in its infancy.}
Open-ended crowdsourcing is presently {\em leveraged
in an unoptimized fashion, with resources wasted and
inaccurate data collected, or even worse, not at all,} leading
to severe impediments to
machine learning. 

\sampar{Challenges.}~The reason why open-ended
crowdsourcing is leveraged in an unoptimized fashion is that 
optimizing open-ended crowdsourcing is substantially harder 
than optimizing boolean crowdsourcing:

\empar{1. Hard to aggregate.}~Due
to the many possible answers to open-ended tasks, 
distilling the `right' answer from this set is non-trivial. 
For example, when drawing a box around an object in an image, 
or transcribing audio into text, 
no two workers will provide the same box or transcription.
This is because the number of possible answers is very large,
and there are many ways of making mistakes. 
In boolean crowdsourcing, we could simply resort to the majority opinion,
but those techniques do not apply, especially when
all of the answers are different from each other.

\empar{2. Sparsity of quality measurements.}
In trying to characterize the error rates of workers,
due to the large number of possible answers, 
it is hard to get reliable estimates of the probability
that a worker provides an answer $a$, when 
the true answer is $b$.
In order to estimate these probabilities, it would take 
a lot of tasks to be issued on crowdsourcing marketplaces
to ensure adequate coverage (and therefore accurate estimates)
for every $(a, b)$ pair.

\empar{3. Many right answers.}~To further complicate matters,
open-ended tasks often have many `right' answers, due to different
underlying perspectives or beliefs, 
making it challenging to distinguish between the case when 
a worker is making mistakes, or the case when the worker 
is simply adopting a different perspective.
For instance, when clustering items---say a 
collection of images of everyday objects, 
workers may use
different criteria---size, color, geometric shape---to 
cluster the items, while also 
inadvertently introducing errors.

\empar{4. Multiple scales.} While boolean crowdsourcing
typically operates on items (images, text) as a whole, open-ended
crowdsourcing can additionally operate on portions of items.
For example, when counting objects in an image, 
we may ask workers to count within portions of 
the image for less error-prone counts.
There are an unbounded number of such portions that can be counted, 
making it hard to pick between them when selecting tasks to
be assigned to workers.

\empar{5. Many open-ended operators.} Unlike boolean crowdsourcing
which is limited to a small number of operator types, 
there is a wide
variety of open-ended operators, 
even for the same problem (including boolean ones as a subset).
For example, for detecting where an object is present in an image, 
workers could draw a box
around an object, fix a box, or compare boxes.
The large number of alternatives makes it hard to design
algorithms.  

\smallskip
In short, these issues (1--3) lead to an increased complexity in reasoning 
about the underlying algorithms, and (4--5) an overwhelming number of
design choices when it comes to designing algorithms.

\sampar{This Paper.}
While there has been a large body of work on open-ended crowdsourcing, primarily
from the HCI (Human-Computer Interaction) community, most of this work has been
on creatively using open-ended crowdsourcing operators in workflows as opposed
to understanding how to model and reason about them, and develop 
optimal algorithms. 
While there has been a deep, interesting, and important body of work from the database
community (and to a certain extent from the AI and machine learning communities)
on optimizing boolean crowdsourcing, 
our hope is to bring open-ended crowdsourcing the same kind of attention,
especially given its importance as articulated above.

Therefore, in this paper, we aim to outline an emerging body of work in optimizing
open-ended crowdsourcing 
from us and from other groups by developing a set of {\em design principles} that we have found to be effective for algorithm development, and by
describing how these design principles were applied for a few papers that
we have been working on in this space.
Note that our survey of the emergent work on open-ended crowdsourcing 
will necessarily be biased by our own work
that we're most familiar with; 
this is not to indicate that the other work is not
as important or as interesting, 
but merely indicates our lack of familiarity with them. 
We will attempt to categorize all of the open-ended crowdsourcing work
from the database community that we are aware of, along with work from
other communities in Section~\ref{sec:related}.

\hidden{
Our aim is to
{\bf \em develop the foundational principles for open-ended crowdsourcing}, i.e.,
\begin{enumerate}
\item {\em formalize and characterize a suite} of open-ended crowdsourcing operators; and
\item use them to develop {\em optimized algorithms} for a set of important open-ended crowdsourcing problems. 
\end{enumerate}
As in our prior work on boolean 
crowdsourcing~\cite{rate,whowon,crowdscreen,searching,humangs,entity-matching-guarantees,entity-matching,finish,pipelines,confidence-general,joglekar2013evaluating,das2016towards}, 
our results will be {\bf \em simultaneously grounded in theory and practice}: this forms the core of our research philosophy. 
If successful, this proposal could represent the next transformative leap 
in our understanding of how to optimally leverage crowdsourcing for data management. 
}

\section{Design Principles for Open-Ended Crowdsourcing}\label{sec:design-principles}
We now describe some approaches that we've found to be
successful in dealing with the increased modeling complexity (issues 1--3 above)
and increased number of design choices (issues 4--5 above), followed
by a solution scaffold or recipe for open-ended crowdsourcing.

\subsection{Dealing with the Modeling Complexity}
The challenges in modeling worker performance stems from 
the fact that there are far too many possible answers that workers
can provide even for a single question or task.
This means that we do not have a 
clear mechanism to aggregate worker answers,
and nor do we have the ability to estimate error rates
of the form $\Pr[\textrm{worker answers $a$}|\textrm{true answer is $b$}]$
for all $(a, b)$ pairs.
We have identified two ways of dealing with this modeling complexity,
both of which essentially allow us to {\em transform}
the worker answer into one or more boolean crowdsourcing answers,
following which we can apply standard techniques from boolean crowdsourcing.

The first approach is to {\em project the answer down} to a finite set of choices.
For example, if the worker provides an answer that is any rational number in a
range, we can project this answer down to a finite set of integers;
yet another way is to project it down into a binary choice: $\leq a$ or $> a$.
Note that this approach is wasteful in that we lose some of
the fine-grained information that workers are providing, and begs
the question of why we didn't simply 
ask the boolean crowdsourcing question
in the first place. 
Nevertheless, this simple approach is commonly used: we provide 
an example of this approach in Section~\ref{sec:counting}.

The second approach is to {\em decompose the answer} down to 
answers to a collection of boolean crowdsourcing questions.
As an example, if a worker is providing a transcription to a piece of 
audio, instead of treating the entire transcription as one open-ended
crowdsourcing task, we can break it down into the 
sequence of individual words, 
each of which can be considered a response or a non-response
to a word transcription task, which is 
much more manageable.
By doing so, we can now model and reason
about workers making mistakes at the level of words, 
rather than complete transcriptions.
In transcription, at least, an additional challenge remains, which is
to identify which words across workers were meant to be 
provided as output for the same portion of the audio piece.
As another example, if a worker is providing a bounding box
as an answer to a detection task, instead of treating
the bounding box as a whole, we can decompose it down into
boolean crowdsourcing answers for individual pixels: 
where if a pixel is part of the box drawn by a worker, the pixel gets a ``yes'' answer
for the corresponding boolean crowdsourcing question,
while it gets a ``no'' answer if it is not.
This approach has the downside that the answers to the 
``pixel-level'' boolean crowdsourcing questions are in fact
not independent---if the answer to a specific pixel is ``yes'',
then it is more likely than the answer to a neighboring 
pixel is ``yes'' rather than no. 
By decomposing the open-ended crowdsourcing task
to boolean crowdsourcing ones, we lose this information.

We also employ a third approach which is fundamentally different from the previous two. This last way of dealing with open-ended crowdsourcing
tasks is a bit more ad-hoc, and problem dependent, 
but does not require us
to discard any information. 
Here, we operate on the answers provided to the open-ended
crowdsourcing tasks directly. 
Consider the answers to a 
single open-ended crowdsourcing task. 
We can represent each of these task answers as nodes in a graph,
and connect nodes that are similar to each other 
(on some similarity metric, such as overlap) 
with an edge annotated by the degree of similarity. 
Once this graph of answers is constructed, we can apply standard graph clustering
algorithms to identify various view-points among
the open-ended task answers:
the largest cluster or clique may represent the ``consensus'' answer.
What can be done with these clusters is dependent on the 
problem. 
We provide an example of this approach in Section~\ref{sec:clustering}.

\subsection{Dealing with the Increased Design Choices}
As described previously, open-ended crowdsourcing
brings with it a considerable increase in the number of 
alternative tasks that can be issued at each step.
The increase is due to the large number of open-ended
crowdsourcing task types available, and also because these
tasks can be applied to portions of items and not just the items directly.
We now describe our approaches to deal with these
increased design choices.

Our first approach is one that has been applied
in the past for boolean crowdsourcing, which is
to estimate the {\em information gain} for issuing a specific
task: here, an additional challenge is that the information gain
may be hard to estimate because we do not have a good
model to reason about worker performance. 
Nevertheless, we may be able to use proxies for information
gain, e.g., prioritizing items or portions of items for which
we have fewer answers than others, or more ambiguity, perhaps
measured by projecting or decomposing the answer down to 
boolean tasks (as described in the previous section).

The second approach is to start at a coarser granularity
and then drill-down only when needed.
For instance, in transcription---we can have workers first 
transcribe the entire audio piece,
and then have additional workers transcribe the portions
that are were found to be ambiguous. 
We will describe another example of this approach 
in Section~\ref{sec:counting}.

A last approach is to incorporate information from primitive 
machine learning algorithms to ``direct'' attention to specific
items or portions of them.
For example, if we know for certain that an object does not
lie in a given portion of the image, we can remove those
portions when providing workers the open-ended task
where they are asked to draw a bounding box about
the object of interest.
Once again, we describe how this approach is used for counting 
in Section~\ref{sec:counting}.

\subsection{Solution Scaffold}
\begin{wrapfigure}{r}{0.6\textwidth}
\centering
\vspace*{-5pt}
\includegraphics[height=1.8in]{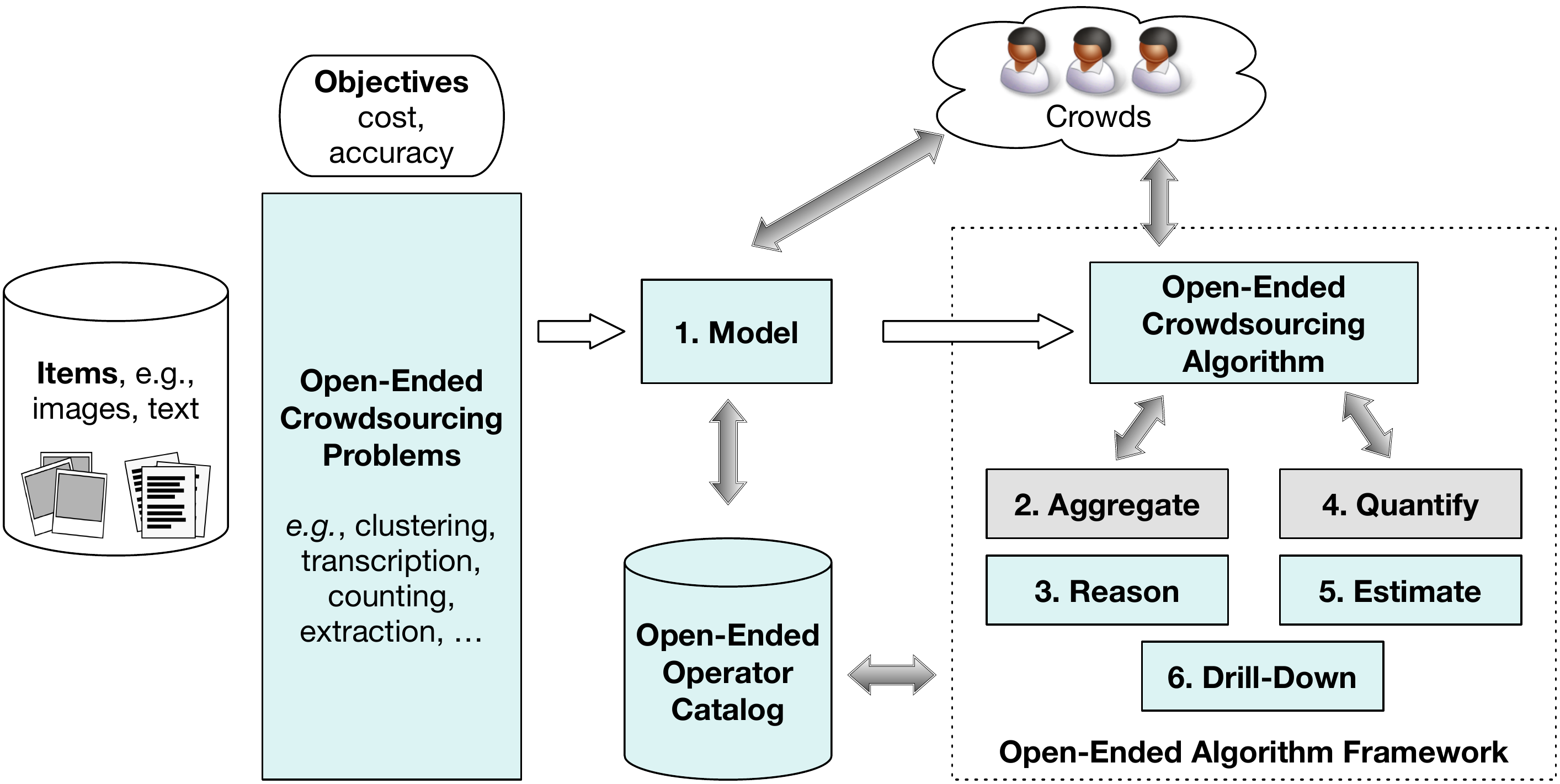}
\vspace*{-5pt}
\caption{\label{fig:flow-v2}\small{Algorithm Flow: Boxes in blue 
did not exist in boolean crowdsourcing, while boxes in gray are substantially different
}}
\vspace*{-5pt}
\end{wrapfigure}

We have developed an {\em open-ended crowdsourcing problem-solving approach}
drawing on our mechanisms to deal with additional
modeling and design choice complexity.
While the specific instantiation may differ across problems,
the overall principles still apply, and the insights transfer across.
For each problem, we apply \mar, short for {\bf \em Model-Aggregate-Reason-Quantify-Estimate-DrillDown}:
the first three (MAR) are tailored to managing modeling complexity,
while the last three (QED) are tailored to managing design choice complexity.

In particular, \mar stands for the following:
(1) {\bf \em Model} the performance of workers on open-ended operators;
(2) Develop methods to {\bf \em Aggregate} across their responses to
identify one or more `right answers';
(3) Identify techniques to {\bf \em Reason} about whether the workers
are generating their answers based on the same or different underlying perspective;
(4) Develop procedures to {\bf \em Quantify} the information gain of different open-ended
operators, and the same operator operating on different items;
(5) Design schemes to incorporate prior {\bf \em Estimates} from automated algorithms
to reduce costs; and
(6) Develop {\bf \em Drill-Down} techniques on the items to enable workers
examine an item more closely.
Then, we {\bf \em design the open-ended crowdsourcing algorithm}
that can leverage (1--6) (if available)---see Figure~\ref{fig:flow-v2}; 
boxes in blue did not exist in boolean crowdsourcing,
while boxes in gray are substantially different. 
Note that it is not necessary to develop solutions for all of (2--6) before
we can reap considerable benefits in practice. 
In the next section, we describe how we applied \mar in practice.

\section{Example Problems in Open-Ended Crowdsourcing}

In this section, we describe our results for two problems that we believe 
are representative of the challenges in open-ended crowdsourcing,
along with our solution approach, in addition to
other problems in this space.

\subsection{Counting}\label{sec:counting}

The goal of {\em counting} is to estimate the number of objects 
of a given type in an image at low cost;
it is a basic computer vision primitive,
with applications in security,
medicine, and biology. 
Counting is a hard problem due to occlusion---the partial 
obscuring of objects behind other objects, 
with state-of-the-art automated techniques 
having accuracies less than 50\%~\cite{zhu2012face,everingham2014pascal}. 

In our paper~\cite{ilprints1128}, we develop cost-effective
techniques to use crowds to accurately count the number of objects 
in images from two completely different domains---cell colonies in microscope photos, 
and people in Flickr photos. 
We now describe the various components of our solution, 
drawing parallels to the \mar methodology.

\emtitle{Model.}
Our open-ended operator is simple---we display an image or a portion
of an image, 
and ask workers to provide the 
count of the number of objects in that portion. 
We find that workers do not make mistakes in counting
when the number of objects in the image is less than a certain number
$k$ ($20$ in our experiments); 
after that, workers start introducing errors, with the errors
growing superlinearly as the number of objects increases.
This may be because workers are not able to keep track
of the objects they have already counted,
or because they get fatigued beyond a certain point
and start introducing errors.
Using this insight, we model worker error
by {\em projecting down} worker answers to boolean ones:
if the answer is $<k$, then it is assumed to be correct;
if the answer is $\geq k$, then it is assumed to be incorrect i.e., 
given a worker provides a count $\geq k$, the only information 
we can deduce is that it is $\geq k$, but no additional information
can be inferred.

While this simple model provides reasonable results,
we are indeed wasting information by ignoring worker answers
if they are $\geq k$. Using a more fine-grained error model
along with maximum-likelihood estimation can help identify a current
best estimate for an image or a portion of an image, allowing us to
``skip ahead'' to the portions that need more attention. That said, this
fine-grained error model requires more expensive training data to estimate
accurately.

\begin{wrapfigure}{l}{0.4\textwidth}
\centering
\vspace*{-10pt}
\includegraphics[height=1.5in]{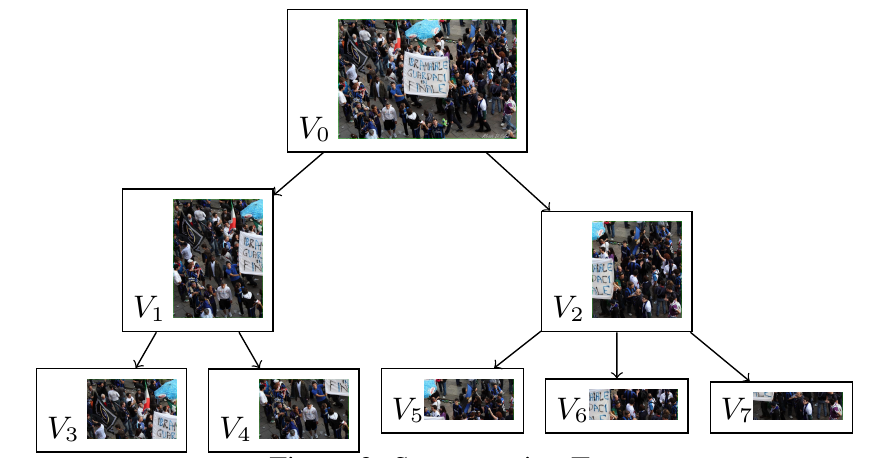}
\vspace*{-22pt}
\caption{\label{fig:segmentation-tree}Segmentation Tree}
\vspace*{-10pt}
\end{wrapfigure}

\emtitle{Drill-down.}
Based on the simple error model,
we can already develop a strategy for counting objects in images,
by repeatedly splitting images and {\em drilling-down}.
We model this process as a {\em segmentation-tree}, 
where the root node represents the original, complete image, 
and children of any node represent the segments obtained by splitting the parent image (using
some splitting scheme, horizontal or vertical). Figure~\ref{fig:segmentation-tree} shows one such example segmentation tree where the original image, $V_0$, is split into segments $\{V_1, V_2\}$, which are respectively split into $\{V_3, V_4\}$ and $\{V_5, V_6, V_7\}$.

We refer to any set of mutually exclusive nodes (i.e. image 
segments) that when put together reconstruct the original 
complete image as a {\em frontier} of the segmentation-tree. 
We start by asking workers to count the number of images in the root node (i.e. the complete image). If they reply with a count greater than or equal to $k$, then we traverse down the segmentation tree by splitting the image and asking workers to provide the count for the children segments. If the count on all segments is smaller than $k$, then we are done. If not, we again split and repeat this process for every segment that still has a count $\geq k$, until we reach a frontier of image segments such that every one of them has a smaller count than $k$.
This simple strategy has optimal {\em competitive ratio} for any given segmentation tree, and also 
returns counts with reasonable accuracies of 93\% when counting people on a standard dataset~\cite{DBLP:journals/corr/PlummerWCCHL15}.
This approach does come with one challenge: objects often span multiple sibling segments in the segmentation tree, and therefore have a danger
of being double-counted. In our paper, we take the easy-way-out by 
asking workers to only count an object if more than half of it appears in the image segment. 
Further improvements may be possible.

\emtitle{Reason and Aggregate.}
Using our simple error model, reasoning about perspectives,
and aggregation 
is not necessary---workers are expected to agree
and provide correct answers as long as the true count is $<k$---thus
we can simply ask one worker to count each image or portion of
the image, and additional answers are not necessary.
In practice, however, we find that sometimes workers may still introduce
errors: to handle this, in our paper, we take three worker answers per 
question and take the median---this simple aggregation scheme is sufficient
to resolve worker differences and obtain high accuracy count estimates. 
Finally, we sum up all the aggregated counts from the different constituent image segments to obtain the count for the original, whole image.
Changes to the worker error model, for instance, using a fine-grained probabilistic model for maximum-likelihood count estimation, will open the road to more interesting aggregation challenges.

\emtitle{Quantify.}
While we did not implement other operators for this problem, there are a number of interesting alternatives that yield more information, at the expense of a little extra worker effort. For instance, we could ask workers to tag objects that they have already counted, say with a dot, to help eliminate double-counting as well as avoid missing objects. Tagged objects could serve as references for other workers, who could then mark additional objects missed by previous workers, or eliminate redundant instances.

\emtitle{Estimate.}
Even if the number of objects is in the hundreds,
our algorithm on the segmentation tree may end up asking several
``useless'' questions at the higher levels all with count $\geq k$, while the ``useful'' questions (whose answers are actually used to compute
the final count at the root) are at the frontier
of the tree where the count is just $k$.
However, it may be possible to use feedback from a primitive
automatic segmentation algorithm to craft a segmentation tree
where there are no useless levels, and where objects do not span 
multiple sibling segments.
Consider the problem of counting cells in
biological images. 
Even though automated counting may be hard due to occlusion,
we can partition the image 
into non-overlapping portions using 
the watershed algorithm~\cite{beucher1992morphological}
and learn prior counts for each partition
using an SVM~\cite{nattkemper2002neural}.
Note that these prior counts may be much smaller
than expected (due to occlusion),
but it suffices as a starting point.
\hidden{This gives us Figure~\ref{fig:bio}(b).}

\hidden{
\begin{wrapfigure}{l}{0.3\textwidth}
\centering
\vspace*{-10pt}
\epsfig{file=img17cells.png,width=0.14\textwidth}
\epsfig{file=img17_Priors.png,width=0.14\textwidth}
\vspace*{-10pt}
\caption{\label{fig:bio}\small{Biological image (a) before and (b) after partitioning}}
\vspace*{-15pt}
\end{wrapfigure}
}

Given these partitions and prior counts,
we can construct a segmentation tree
that groups multiple contiguous partitions together
until they hit up to $k$ objects each 
(based on the prior counts,
which may be underestimating).
This allows us to construct a segmentation tree
where all levels are ``useful'' given our prior information. 
Since merging partitions together optimally is {\sc NP-Hard}
via a reduction from planar partitioning~\cite{dyer1985complexity},
we employ other heuristic techniques in our paper,
such as first-fit~\cite{coffman1996approximation}.
We then traverse the tree asking workers as before.
By using the prior machine-learned estimates in this fashion, we are able to skip several ``useless'' questions providing a 2x reduction in cost. It should be noted that the images of cell colonies are much more amenable to automated prior estimation techniques. It is an open challenge to explore whether similar techniques could be applied to other kinds of objects, where it is a lot harder to get accurate prior counts.

\subsection{Clustering}\label{sec:clustering}


Given a collection of items (e.g., images, documents),
the goal of clustering, is to organize them into coherent groups.
Collections of images or documents are commonplace;
organizing them is essential before one can understand
the themes in the collection, or improve search or browsing.
Clustering embodies all the challenges faced by open-ended crowdsourcing: 
workers can have different perspectives leading to multiple ``right'' answers making 
worker responses hard to aggregate. 
The space of possible responses is also extremely large, since workers operate on multiple items simultaneously. 
The open-endedness of the problem also means that 
there are a number of different interfaces and operators 
that could conceivably be used. 
We shall now see how applying the \mar approach allows us to reason about this challenging open problem in a principled fashion.

\emtitle{Model.}
Prior work has considered crowdsourced clustering, limited to the boolean operator
where pairs of items are compared~\cite{crowdclustering,Yi12crowdclusteringwith}---as a result, crowd workers do not have any context to compare items. 
Also, the eventual clusterings end up having ``mixed'' perspectives,
resulting in low accuracies.  
Instead we used a basic open-ended clustering operator, 
where a set of items
are provided to workers, and they are asked to group 
them into an arbitrary number of disjoint clusters.
Using this operator, we had multiple workers cluster a stylized image dataset 
with each image containing a shape with different sizes and colors, as a running example. 
First, we introduce the notion of {\em concept hierarchies} to 
capture the notion of worker cluster perspectives.
One concept hierarchy for the concept of shapes, 
could be to have \{{\em All Shapes}\} divided into \{{\em Quadrilaterals}, 
{\em non-Quadrilaterals}\}, 
the latter of which is subdivided into \{{\em Triangles}, {\em Circular Shapes}\}. 
For any given dataset, the concept hierarchy need not be unique. 
For instance, another hierarchy would have  \{{\em All Shapes}\} divided into
\{{\em Circular Shapes}, {\em Straight-Edged Shapes}\}.
When clustering, each worker answer can be seen to draw from one or more inherent concept hierarchies.

\begin{wrapfigure}{l}{0.4\textwidth}
\centering
\vspace*{-10pt}
\epsfig{file=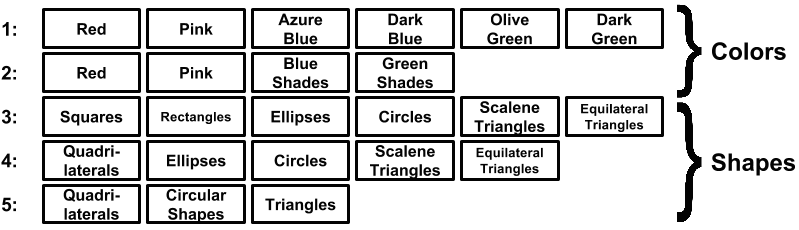,width=0.4\textwidth}
\vspace*{-20pt}
\caption{\label{fig:worker-clusterings} Examples of real worker clusterings}
\vspace*{-12pt}
\end{wrapfigure}

Figure~\ref{fig:worker-clusterings} shows examples of real worker clusterings on our dataset. While workers 1 and 2 clustered based on color alone, workers 3, 4 and 5 clustered based on shape. We focus on the latter for the time being. 
One conceptual hierarchy, $C$, ``consistent'' with workers 3, 4, and 5 is \{{\em All Shapes}\} divided into \{{\em Quadrilaterals}, {\em Triangles}, {\em Circular Shapes}\}, which are respectively subdivided into \{{\em Squares}, {\em Rectangles}\}, \{{\em Scalene Triangles}, {\em Equilateral Triangles}\}, and \{{\em Circles}, {\em Ellipses}\}.
We additionally introduce the notion of {\em frontiers} on a given concept hierarchy to capture the notion of granularities: a frontier in a hierarchy is a set of nodes that do not have any ancestor-descendent relationship between them, and together cover all paths to the leaves. 
Each frontier corresponds to one valid granularity of clustering 
consistent with the concept hierarchy. 
Representing the concept hierarchy, $C$, as a tree, we have worker 3 operating at the leaf nodes, or at the finest granularity of the tree, corresponding to the frontier \{{\em Squares}, {\em Rectangles}, {\em Scalene Triangles}, {\em Equilateral Triangles}, {\em Circles}, {\em Ellipses}\}. Similarly, worker 5 is operating at a depth of one in the tree, which is the coarsest non-trivial granularity and corresponds to the frontier \{{\em Quadrilaterals}, {\em Triangles}, {\em Circular Shapes}\}.

\emtitle{Reasoning and Aggregation.}
From our experiments on the stylized dataset, 
we make the following observations: 
(a) Workers cluster using different {\em perspectives},
e.g., some workers clustered using shape, and others clustered using color.
That said, there was a dominant, popular clustering perspective,
in this case, shape. 
(b) Even within a perspective, workers cluster
at various {\em granularities}, e.g., 
some workers clustered shapes into rectangles
and non-rectangles, while others broke up non-rectangles
into fine-grained clusters.
(c) Sometimes, there are {\em confusing items} that end up 
being placed in different clusters by different workers
even if they agree on the perspective and the granularity.

To reason about worker's perspectives, 
we can develop a notion of {\em consistency}:
two worker clusterings (i.e., a set of clusters formed by each worker) 
are consistent if for any two pairs of clusters $C, C'$,
one from each worker, either $C \subseteq C', C' \subseteq C,$ or $C \cap C' = \emptyset$,
i.e., each cluster from one worker generalizes, specializes, or does not overlap with another worker's clusters.
This definition allows consistent workers to cluster at different granularities.

Given the notion of consistency, we can now directly operate on worker responses to identify
whether there are any ``consensus'' clusterings, i.e., consensus concept hierarchies (perspectives), and frontiers within them (granularities), that emerge. 
This allows us to have a starting point to cluster the rest of the items.
To do this, we generate a {\em clustering graph}, with one node per worker, 
with consistent pairs having an edge between them. 
\hidden{As we have noted, workers cluster items on multiple valid perspectives, none of which can be called better or more correct than others. This begs the question: how do we formally reason about these different perspectives in order to aggregate worker responses? What is the underlying ``true'' clustering that we wish to converge upon?
We developed techniques to map worker clusterings to hierarchies (to identify worker perspectives), and formalize the idea of the true underlying {\em consensus} as the {\em maximum likelihood hierarchy}. Our formalization is built on the idea of consistency, and captures the idea that even differing clusterings can be resolved, as long as the underlying concept hierarchies are consistent with each other. 
An important feature of the consensus hierarchy is that it captures all possible granularities on which a worker may cluster the given items through the notion of frontiers. In our paper, we also solve the problem of finding the maximum likelihood frontier from the consensus hierarchy.

Consider the {\em clustering graph} formed with one node per worker, with consistent
pairs having an edge between them. 
}
\begin{wrapfigure}{l}{0.2\textwidth}
\centering
\vspace*{-10pt}
\epsfig{file=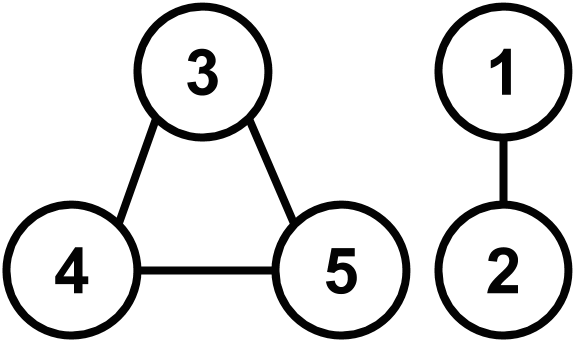,width=0.17\textwidth}
\vspace*{-10pt}
\caption{\label{fig:clustering-consistency-graph} Clustering graph}
\vspace*{-10pt}
\end{wrapfigure}
Figure~\ref{fig:clustering-consistency-graph} shows the clustering graph corresponding to the workers from Figure~\ref{fig:worker-clusterings}. Workers 1 and 2 both clustered based on color alone and were consistent with each other, while workers 3, 4, and 5 clustered by shape alone with no inconsistencies among themselves.
We can show that under multinomial worker perspective selection
models, the maximum likelihood worker perspective
is the {\sc Max-Clique} in the graph. 
Despite {\sc Max-Clique} being {\sc NP-Hard},
the clustering graph is often not large, making the problem
tractable. In our paper, we additionally describe how we can incorporate
worker error models, which introduce additional complexity to the problem.

\hidden{
We have not yet considered the possibility of worker errors----aggregating potentially erroneous clusterings is another interesting aggregation problem. In our paper, we reason about worker errors in terms of disagreements between them. The problem of aggregation with errors can then be mapped into the problem of finding items that workers disagree on, and removing them to converge on consistent clusterings. However, we show that even finding the smallest number of items
to remove to make two worker clusterings consistent
is {\sc NP-Hard}, since it is equivalent to the problem of 
of finding the smallest {\sc Dominating Set} in bipartite graphs~\cite{liedloff2008finding}.
As next steps, we are exploring fixed parameter tractable algorithms to solve this problem,
specifically via a kernelization technique,
motivated by a similar application for {\sc Cluster Vertex Deletion}~\cite{boral2016fast}. 
}

\emtitle{Quantify and Drill-Down.}
So far, we have described how to use a single open-ended clustering operator and
aggregate responses from it for a small number of items.
However, if we have a large number of items, workers might not be able to
cluster all of them in one go.
This suggests the need for drilling-down, or splitting the set of items, and asking workers to cluster the resulting subsets. This raises the challenge of aggregating partial clusterings. 
For this purpose, we maintain a {\em kernel} of items across clustering tasks, 
akin to {\em pivots} in Polychronopoulos et al.~\cite{DBLP:conf/webdb/PolychronopoulosADGP13},  
to be able to relate partial clusterings across each other, for aggregation.
We design techniques to extend the current maximum likelihood hierarchy by merging worker responses on new items to the existing hierarchy---we leverage these techniques to design a merging algorithm which aggregates clusterings from separate subsets together to output a consensus hierarchy on the original, complete set of items.

We additionally incorporate a categorization-based operator,
once the consensus clustering is identified, to provide additional cost benefits,
by categorizing the remaining items into the discovered clusters~\cite{humangs}.
Instead of having workers repeatedly cluster many items (implicitly a many-to-many comparison),
they only need to identify which cluster or category is appropriate for one item 
at a time, with the clusters being fixed. 

\hidden{
While we have studied one natural open-ended clustering operator in our work, there are other operators worth exploring. One option in particular that we propose in our paper is to use a new, 
categorization-oriented operator, in conjunction with the existing clustering operator.
Under this operator, workers are shown some existing clusters,
each depicted using a small number of example items, 
and are asked to put a new item into one of the shown clusters. 

\begin{wrapfigure}{r}{0.3\textwidth}
\vspace{-10pt}
\scriptsize
\begin{tabular}{ |c|c|c|c|c| }
\hline
& \multicolumn{2}{ |c| }{\textsf{scenes}~\cite{fei2005bayesian}} & \multicolumn{2}{ |c| }{\textsf{imagenet}~\cite{deng2009imagenet}} \\
\hline
 &  Re & Ac & Re & Ac \\
\hline
Ours &  \textbf{0.985} & \textbf{0.992}  &  \textbf{0.800} & \textbf{0.881} \\
\hline
\cite{crowdclustering} & 0.161 & 0.319 & 0.241 & 0.465 \\
\hline
\cite{Yi12crowdclusteringwith} &  0.089 & 0.226 & 0.268 & 0.457 \\
\hline
\end{tabular}
\vspace{-10pt}
\caption{\label{fig:clust-results}Preliminary Comparison}
\vspace{-15pt}
\end{wrapfigure}
} 

\hidden{
While in our paper we have not explicitly considered using prior estimates in our various clustering algorithms, there are a number of interesting ways in which to incorporate apriori information into our algorithm design. For instance, given some prior estimate of the true clustering hierarchy, we can pick the kernel intelligently, so as to minimize the overlap of clusters from different subsets of items. We could also such knowledge trade off clustering and categorization optimally.
It is also conceivable that we can obtain similarity estimates on pairs of items using automated learning algorithms, and use these estimates to streamline our clustering process---for instance, while drilling-down and separating items into subsets for clustering, it may improve worker performance and information gain if each subset has some potentially ``similar'' items for context, as well as some ``tricky'' pairs for disambiguation.
} 

Overall, our techniques lead to up to 3$\times$ better recall and 1.9$\times$ better accuracy than boolean clustering schemes using pairwise judgments~\cite{crowdclustering,Yi12crowdclusteringwith} on their datasets, 
for the same crowdsourcing cost.

\subsection{Other Problems}

We've applied the \mar approach to other open-ended crowdsourcing problems.
We describe some of these problems briefly here, followed by work done by others.

\sampar{Extraction.} 
The goal of {\em extraction} is to use crowdsourcing to gather 
entities of a specific type~\cite{Trushkowsky:2013kh}.
We considered a broad space of open-ended operators for this problem,
leveraging the 
attributes of the entity set~\cite{2015arXiv150206823R}.
For example, musicians in Chicago can be categorized
as guitarists, drummers, and so on, and may play music 
of various types.
By considering these attributes, we can ask workers
to answer more {\em fine-grained} open-ended questions:
e.g., provide a jazz drummer in Chicago. 
This allows us to target the questions at the attribute combinations
where we lack entities. 
However, the number of possible open-ended questions
increases exponentially in the attributes, making the problem challenging.
\hidden{To mitigate this, we exploit correlations across
the answers to the open-ended questions:
the results corresponding to ``provide a drummer''
can help judge whether ``provide a country drummer''
will help.} 
We showed that picking the best questions is
{\sc NP-Hard}, and found that a {\em drill-down} based technique works well,
where we ask generic questions first (e.g., provide a musician), and then drill down and ask more specific questions later (e.g., provide a drummer in Lincoln Park). 

\sampar{Searching.}
The goal of 
crowd-assisted search~\cite{datasift,datasift-demo}, 
is to return relevant results from a corpus 
given a search query with embedded images, video, and/or text.
Here, we once again found that the following open-ended problem solving aspects are helpful:  (a) multiple open-ended operators; (b) starting at a coarser granularity and drilling down
into more promising candidates; and 
(c)  incorporating prior information -- in this case from regular text search engines.

\sampar{Batching.}
Boolean tasks are typically grouped into batches of 10s to 100s of
tasks that are attempted by workers together---to reduce cost and effort---making it essentially
one large open-ended task. 
However, these tasks are assumed to be answered
independently, which is not the case.
We developed a probabilistic model to reason about the answers
incorporating two forms of judging:
independent, where workers answer each question
independently, or correlated, where 
workers implicitly rank the items and then pick the 
top-$k$ to be positive, with the rest negative (i.e., the Plackett Luce model). 
We found that this model, 
let to substantial accuracy improvements over schemes
that ignored these correlations~\cite{DBLP:conf/kdd/ZhuangPRH15}.

\smallskip
\noindent In addition, there has been work by others on 
optimizing other open-ended crowdsourcing problems.
We describe some of them below. 

The goal of {\em transcription} is to transcribe a sequence of 
words into text; it is a very challenging problem, 
with automated techniques performing very poorly~\cite{yu2012automatic,lasecki2012real}.
By decomposing worker answers down into individual words, one can 
apply standard multiple sequence alignment algorithms from the bioinformatics literature 
to align worker answers and identify consensus answers, leading to substantial improvements~\cite{lasecki2012real}; other work tries combining crowd answers with automated techniques~\cite{williams2011crowd,van2015improving}.
Some prior work has also looked at the problem of {\em detection} --- finding the location of objects in images, applying computer vision models and human input via a Markov Decision Process~\cite{russakovsky2015best}.

Other open-ended crowdsourcing work exists as well, for various purposes,
including designing plans~\cite{DBLP:conf/icde/LotoshMN13},
rule mining~\cite{DBLP:conf/sigmod/AmsterdamerGMS13}, 
and pattern matching~\cite{DBLP:conf/cidr/DemartiniTKF13}.

\section{Related Work}


\label{sec:related}
The work on open-ended crowdsourcing is related to work in many areas.

\stitle{Area 1: Optimized Boolean Crowdsourcing.}
There has been quite a bit of work on optimizing boolean crowdsourcing,
targeting basic algorithms, 
including filtering~\cite{crowdscreen,rate,finish},
sorting~\cite{markus-sorts-joins},
max~\cite{venetis1,DBLP:conf/www/VenetisGHP12, whowon}, categorization~\cite{humangs}, 
top-K~\cite{DBLP:conf/webdb/PolychronopoulosADGP13,DBLP:conf/icdt/DavidsonKMR13, searching}, spatial crowdsourcing~\cite{DBLP:journals/geoinformatica/DengSDZ16,DBLP:journals/tsas/ToSK15},
and entity resolution (ER)~\cite{crowder,Demartini:2012bt,chu2015katara,wang2013leveraging,Gokhale:2014wv,tong2014crowdcleaner,wang2015crowd, entity-matching-guarantees,entity-matching}.

\stitle{Area 2: Crowdsourcing Systems and Toolkits.}
Many groups have been building crowdsourcing systems and toolkits to harness
crowdsourcing in a ``declarative'' manner~\cite{crowddb, qurkCIDR,deco,deco-qp},
as well as several domain-specific toolkits~\cite{DBLP:conf/www/BozzonBC12,tova-trivia,datasift,datasift-demo}.
All these systems and toolkits could benefit from the design of optimized  algorithms
as building blocks.

\stitle{Area 3: Quality Estimation:} A number of papers perform simultaneous worker quality estimation and most accurate answer estimation, typically using the EM algorithm,
sometimes providing probabilistic or partial guarantees, 
and sometimes modeling difficulty, bias, and adversarial behavior, e.g.,~\cite{karger, whitehill-accuracy, cdas, dalvi2013aggregating, zhang2014spectral,swiftly, joglekar2013evaluating,confidence-general, das2016towards}.
To our knowledge, there is no work on applying EM
to open-ended crowdsourcing tasks.

\stitle{Area 4: Decision Theory:} Recent work has leveraged decision theory 
for improving cost and quality in simple workflows, 
typically using POMDPs (Partially Observable MDPs),
to dynamically choose the best decision to make at any step, e.g.. ~\cite{DBLP:conf/uai/LinMW12,DBLP:conf/aamas/KamarHH12}. 
While some of this work could be applicable to some open-ended
tasks, there are no optimality guarantees associated with any of
these techniques.

\section{Conclusion}
Open ended crowdsourcing is not only challenging and interesting, but also necessary to meet the demands of the new generation of data-hungry applications. We hope that the next wave of crowdsourcing research from the database community will tackle more problems in this space, expanding the frontier of our understanding.


\label{sec:conclusions}

{\scriptsize
\bibliographystyle{ieeetr}
\bibliography{ref}
}

\end{document}